\documentclass[aps,pre,twocolumn,superscriptaddress,nofootinbib,floatfix,showpacs]{revtex4-1}
\usepackage{epsfig}
\usepackage{amsmath,amssymb,amsfonts}
\usepackage{hyperref}
\usepackage{mathrsfs}
\usepackage{bbm}
\usepackage{slashed}
\usepackage{graphicx}
\usepackage{verbatim}
\usepackage{url}
\usepackage[normalem]{ulem}

\usepackage[usenames]{color}

\newcommand{\be}{\begin{equation}}
\newcommand{\ee}{\end{equation}}

\newcommand{\bi}{\begin{itemize}}
\newcommand{\ei}{\end{itemize}}
\newcommand{\bea}{\begin{eqnarray}}
\newcommand{\eea}{\end{eqnarray}}

\newcommand{\mb}{\mathbf}

\newcommand{\RefA}[1]{{\textcolor{black}{#1}}}  
\newcommand{\RefB}[1]{{\textcolor{black}{#1}}}  
\definecolor{darkgreen}{rgb}{0.1,0.7,0}
\definecolor{bensblue}{rgb}{0.1,0.1,0.9}

\usepackage[T1]{fontenc} \usepackage[latin1]{inputenc}

\begin{document}
%
\title{\textit{In situ} characterization of ultraintense laser pulses}

\author{C.~N.~Harvey}
\email[]{cnharvey@physics.org}
\affiliation{Department of Applied Physics, Chalmers University of Technology, SE-41296 Gothenburg, Sweden}

\begin{abstract}
We present a method for determining the characteristics of an intense laser pulse by probing it with a relativistic electron beam.  After an initial burst of very high-energy $\gamma$-radiation the electrons proceed to emit a series of attosecond duration X-ray pulses as they leave the field.
These flashes provide detailed information about the interaction, allowing us to determine properties of the laser pulse: something that is currently a challenge for ultra-high intensity laser systems.
\end{abstract}
\maketitle

\section{Introduction}
During recent decades there has been an exponential increase in the powers and intensities of state of the art laser facilities \cite{PhysRevSTAB.5.031301}. Peak focal intensities of the order of $10^{22}$ Wcm$^{-2}$ can now be achieved in the laboratory \cite{Yanovsky:2008} and this is expected to be exceeded by at least an order of magnitude as new facilities come online. These facilities, which include the Vulcan 20 PW upgrade \cite{Vulcan}, the Extreme Light Infrastructure (ELI) Facility \cite{ELI} and the XCELS project \cite{XCELS}, have stimulated a large body of research in classical and quantum strong field physics, overviews of which can be found in Refs.~\cite{Heinzl:2011ur, DiPiazza:2012RevModPhys, turcu2016high}.

Despite the promise of such high-intensity fields, determining the exact properties of an intense laser pulse created in the lab remains a significant challenge. \RefA{While standard optical metrology can be carried out while running the laser at lower power, this is doesn't necessarily give an accurate representation of the pulse at higher intensities.} Without detailed information regarding the pulse's peak intensity, duration and polarisation the planning and execution of experiments becomes difficult. 
\RefA{One avenue of research is in the multiple ionisation of different atomic species in the laser focus to determine the peak intensity, however this requires a detailed understanding of time-dependent ionisation cross-sections for a variety of atomic species \cite{Popruzhenko}.}
Several mechanisms have been proposed to \RefA{directly} extract information about high-intensity laser pulses (e.g. of the range $\sim 10^{21}-10^{22}$ Wcm$^{-2}$), such as peak intensity \cite{Har-Shemesh:12}, carrier envelope phase \cite{PhysRevLett.105.063903, PhysRevE.85.035401}, and duration \cite{mackenroth2017determining} based on the radiation emitted by electrons subjected to such fields.
\RefA{The advantage of this type of metrology is that the pulse properties are measured under the same conditions as present in the subsequent experiments.}

In this paper we propose a method valid up to the case of extreme intensity fields (i.e.~$\gtrsim10^{23}$ Wcm$^{-2}$) where energy losses due to radiation emissions influence the dynamics of electrons inside the pulse. 
Most literature on this topic is concerned with what happens when electrons first enter such fields. It is at this point that radiation emissions are strongest and deceleration most violent.
Such radiation reaction (RR) effects cause the electrons to lose most of their energy before they reach the peak of the pulse \cite{PhysRevA.96.022128, PhysRevA.90.053847}, meaning that collisions typically end with lower energy (but still relativistic) electrons meeting the most intense part of the field. The radiation emissions at this point are of much lower energy (10-100's KeV) and so typically overlooked in the backdrop of the much higher (10-100's MeV) emissions driven by RR at the start of the collision. (One notable exception is the recent proposal to use such radiation to determine the carrier envelope phase of an intense field \cite{LiPRL}.)
However, although the most significant (longitudinal) acceleration is over, upon nearing the peak of the pulse the electrons are accelerated around strongly in the transverse direction, emitting a series of short bursts of radiation in time with the rise and fall of the field. By measuring the angles and amplitudes of these femto-second duration pulses we show that it is possible to determine the intensity, duration and polarisation of the laser field. 

\section{Theory}
Adopting natural units where $\hbar=c=1$ we start by considering a plane wave field propagating in the $z$ direction described by the null wave vector $k^\mu=\omega_0 n^\mu=\omega_0 (1,0,0,1)$, with central frequency $\omega_0$ (in a later section we will progress to focussed fields). The field can be polarised in both the perpendicular directions, the degree of which is quantified by the two polarisation vectors
$\varepsilon_x=(0,\delta_x,0,0)$,
$\varepsilon_x=(0,0,\delta_y,0)$,
where $\delta_x=1/\sqrt{1+\delta^2}$ and $\delta_y=\delta/\sqrt{1+\delta^2}$, so that $\delta\in [0,1]$ defines the polarisation, with $\delta=0$ being linear and $\delta=1$ being circular. These basis vectors satisfy $k^2=k\cdot\varepsilon_{x,y}=0$, $\varepsilon_{x,y}^2<1$, and we use them to construct an electromagnetic field tensor describing the laser pulse
\begin{equation}
F^{\mu\nu}(\phi)=a_0 (f_x(\phi)f_x^{\mu\nu}+f_y(\phi)f_y^{\mu\nu}),
\end{equation}
where $f_{x,y}^{\mu\nu}\equiv n^\mu\varepsilon_{x,y}^\nu-n^\nu\varepsilon_{x,y}^\mu$, and $f_{x,y}(\phi)$, satisfying $f_{x,y}(-\infty)=f_{x,y}(\infty)=0$ is a function describing the pulse shape profile, in this case taken to be a Gaussian envelope.
The field tensor is taken to depend on the phase $\phi\equiv k\cdot x=\omega_0(t-z)$, and we have introduced a dimensionless measure of peak field intensity defined in the usual manner: $a_0\equiv eE/m\omega_0$, where $e$ is the electron charge and $m$ the mass. We take this opportunity to similarly define a time dependent measure of intensity $a=a(\phi)=\sqrt{a_0^2(f_x^2(\phi)+f_y^2(\phi))}$, such that $a=a_0$ at the peak of the field.

The motion of an electron in such a field would ordinarily be governed by the Lorentz equation, but in cases of high intensity the strong acceleration gradients result in significant emissions of radiation causing the particle to lose energy. These RR effects are taken into account via a correction term to the Lorentz equation. However, determining the correct form of this term is surprisingly non-trivial. 
Here we adopt the perturbative approach of Landau and Lifshitz \cite{LLII} where the second derivate of the four-velocity is approximated using the Lorentz term.  Then the equation of motion is given by 
\begin{eqnarray}
 \dot{u}^\mu = &&\frac{e}{m} F^{\mu\nu} u_\nu + \frac{2}{3} r_e \bigg\{
  \frac{e}{m^2} \dot{F}^{\mu\nu} u_\nu + \nonumber\\
 && \frac{e^2}{m^3}
  F^{\mu\alpha}F_\alpha^{\;\;\nu} u_\nu - \frac{e^2}{m^3} u_\alpha
  F^{\alpha\nu}F_\nu^{\;\;\beta} u_\beta \, u^\mu \bigg\},\label{LL1}
\end{eqnarray}
where $r_e=e^2/4\pi m$ is the classical electron radius, and $u^\mu$ the four-velocity.
Equation (\ref{LL1}) is valid when the radiation reaction force is much smaller than the Lorentz force in the instantaneous rest frame of the particle.  There exist alternative equations in the literature (for an overview see \cite{Burton:2014,Vranic:2015}) and, while it is still an open problem as to which is the correct formulation,  the Landau-Lifshitz equation is consistent with quantum electrodynamics (QED) to the order of the fine-structure constant $\alpha$ \cite{0038-5670-34-3-A04,Ilderton:2013tb}.

It is instructive to provide an estimate for when RR effects become important.  Using just the Lorentz force to determine the motion, the radiated power $P$ is given by Larmor's formula in terms of the instantaneous acceleration,
\begin{eqnarray}
P=\frac{2}{3}m r_e \textrm{acc}^2=\frac{2}{3} r_e m\omega_0^2 a^2 \gamma (1+\beta),
\end{eqnarray}
Normalizing this by $\omega m$ we obtain the energy loss per cycle in terms of the electron rest energy $mc^2$ \cite{Koga:2005, Harvey:worldline}
\begin{eqnarray}
R\equiv \frac{P}{\omega_0 mc^2}=\frac{2}{3}r_e\omega_0 a^2\gamma (1+\beta).\label{R}
\end{eqnarray}
When this parameter reaches unity we are in the ``radiation-dominated regime'' \cite{PhysRevLett.102.254802}, where RR effects are of the same magnitude as the Lorentz force \footnote{Note that the Landau Lifshitz equation is still valid in this regime since we are not referring to the particle rest frame.}.   

Additionally, we must distinguish between regimes where classical RR effects dominate and where QED effects become important. With this in mind we introduce the invariant ``quantum efficiency parameter'' $\chi=\sqrt{p^\mu F_{\mu\nu} p^\nu}/m^2 \approx a\omega_0\gamma(1+\beta )/m \sim \gamma E/E_{\textrm{cr}}$, where $E_{\textrm{cr}}=1.3\times10^{16}$ Vcm$^{-1}$ is the QED ``critical'' field (``Sauter-Schwinger'' field) \cite{QEDcriticalfield1,QEDcriticalfield2,QEDcriticalfield3}.  The parameter $\chi$ can be interpreted as the work done on the particle by the laser field over a Compton wavelength. In the regime $\chi\sim1$ quantum effects will dominate. We find that in the region of interaction most of interest $R\lesssim 1$, while $\chi\ll 1$ and so we simulate our setup classically.  Modelling using stochastic QED routines is more relevant to cases where we are interested in the effect of small numbers of high energy photons \cite{PhysRevE.92.023305, blackburn2018benchmarking}. In our study the region of interest is when RR is dominated by the effects of large numbers of low energy emissions making a classical model more appropriate.

Once we have calculated the particle trajectory, the resulting radiation emissions can be obtained via the Li\'enard-Wiechart potentials. Deriving an expression for the energy radiated per unit solid angle per unit frequency one finds \cite{jackson},
\be 
\frac{d^2I}{d\omega^\prime d\Omega} 
= \left|\int\limits_{-\infty}^{\infty}\frac{\mb{n}\times[(\mb{n}-{\beta})\times\dot{\mb{\beta}}]}{(1 - \beta\cdot \mb{n})^2}       
e^{i\omega^\prime/\omega_0 [t + D(t)]} dt \right|^2, 
\label{spectrum}
\ee
where $\mb{n}$ is a unit vector pointing from the particle's position to a detector ($D$) located far away from the interaction, and $\beta$ and $\dot{\beta}$ are, respectively, the particle's relativistic velocity  and acceleration. Here  we have normalized the intensity by the factor $e^2/4\pi^2 $. All the quantities in the above equations are evaluated at the retarded time so one can directly do the integration in some finite limit. 

In the case of high-intensity fields Eq.~(\ref{spectrum}) becomes very computationally expensive to evaluate since it involves quadrature over highly oscillatory functions \cite{PhysRevSTAB.13.020702}. An alternative method is to calculate the emission spectra using a novel Monte Carlo method introduced in \cite{Wallin:2014moa}.
The method has been incorporated into the code SIMLA \cite{Green:2014kfa} and works as follows.
The particles in the simulation are relativistic which means the radiation due to transverse acceleration is dominant, since this is a factor $\gamma^2$ larger than that due to longitudinal acceleration \cite{jackson_classical_1999}.
Since the acceleration and velocity of the particle are perpendicular, the radiation can be approximated as synchrotron radiation.
To do this we calculate the \emph{effective magnetic field}, $H_{\text{eff}}$, acting on the particle over each timestep in the simulation, i.e.~the magnetic field which would cause the same acceleration as the electric and magnetic fields together.
The representative frequency of synchrotron emission can then be expressed as $\omega_c=3eH_{\text{eff}}\gamma^2/2m$. 
For a relativistic particle in an external, homogenous magnetic field, 
the classical radiation cross section can be expressed in terms of the intensity given by \cite{jackson_classical_1999}
\begin{equation}
	\frac{ \partial \Gamma_\textrm{cl}}{\partial \omega^\prime} = \frac{1}{\omega^\prime} \frac{ \partial I}{\partial \omega^\prime} = \frac{\sqrt{3}}{2 \pi} \frac{e^3 H_\textrm{eff}}{\omega^\prime m} F_1(\omega^\prime/\omega_c),\label{Erikclassical}
\end{equation}
where
$F_1(\xi) = \xi \int_{\xi}^{\infty} K_{5/3}(\xi^\prime) \mathrm{d}\xi^\prime$ is the first synchrotron function.
(We note that (\ref{Erikclassical}) is integrable in the limit $\omega^\prime\rightarrow 0$ and therefore the expression is well-defined. For further details see Ref.~\cite{PhysRevLett.89.094801}.)
At each timestep in the code we calculate $\omega_c$ and then use a Monte-Carlo method to sample from the spectra. Once we have the emission frequency, the direction of emission is taken to be that of the particle velocity, which is a good approximation for relativistic particles \cite{jackson_classical_1999}. To remove all doubt, we have calculated a number of different cases over the full range of parameters we are considering using the Li\'enard-Wiechart method (\ref{spectrum}) and found it to be in excellent agreement with the method presented here.

\section{Results}
\begin{figure}
\includegraphics[width=1.0\columnwidth,]{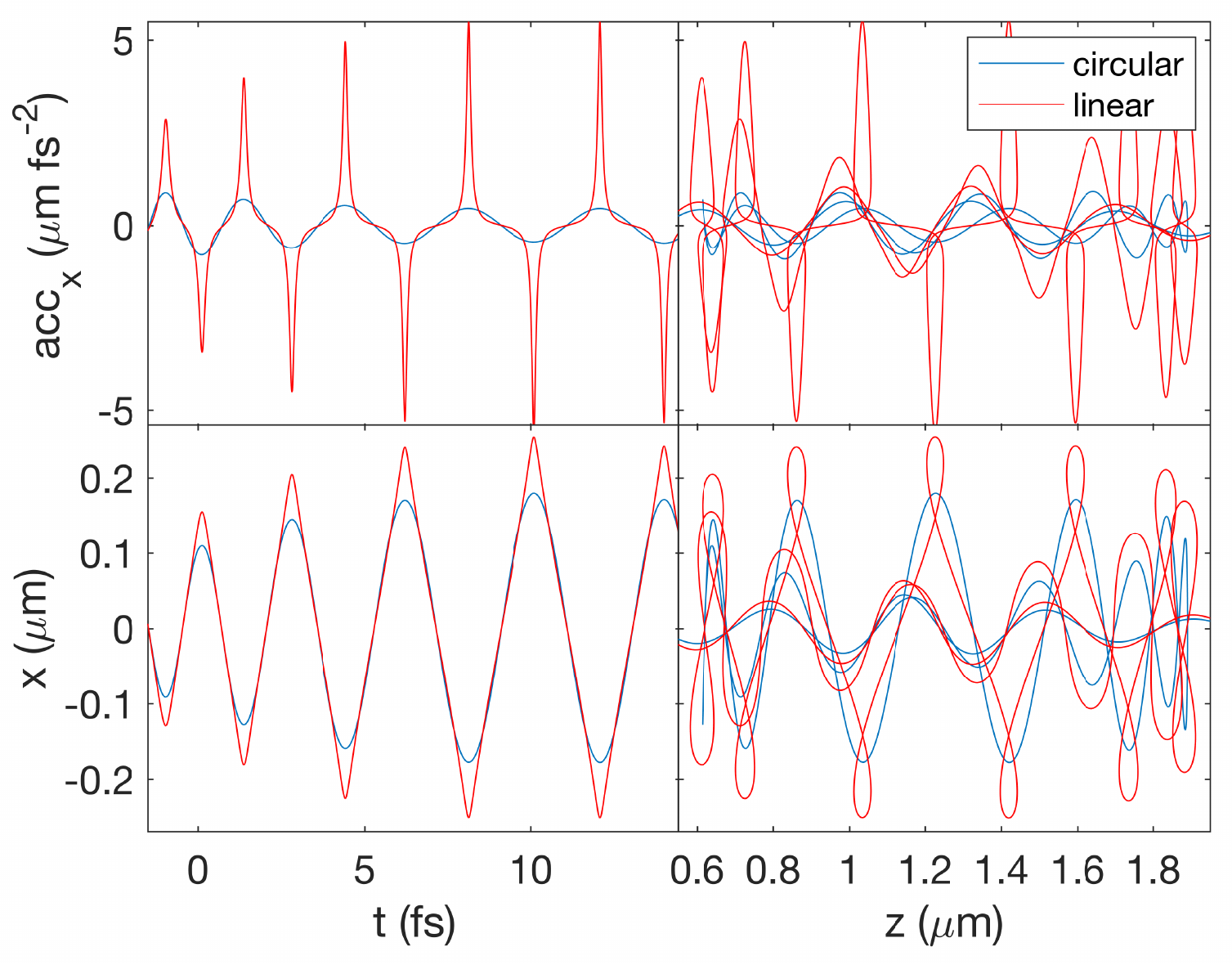}
\caption{Lab-frame trajectories and accelerations of an electron in linear and circularly polarised laser pulses. The electron has an initial $\gamma_0=1000$  and collides with a 27fs (10-cycle) FWHM laser pulse of peak intensity $a_0=200$ and wavelength 800nm.\label{fig:traj} }
\end{figure}

We consider the collision between a relativistic electron and an intense laser pulse.
For the purposes of comparison we define a baseline configuration where the electron has an initial $\gamma_0=1000$ (511 MeV) and is brought into collision with a laser pulse of wavelength 800nm, peak intensity $a_0=200$ ($3.4\times 10^{23}$Wcm$^{-2}$) and of duration 27fs FWHM (i.e. 10 cycles).

\begin{figure}
\includegraphics[width=1.0\columnwidth,]{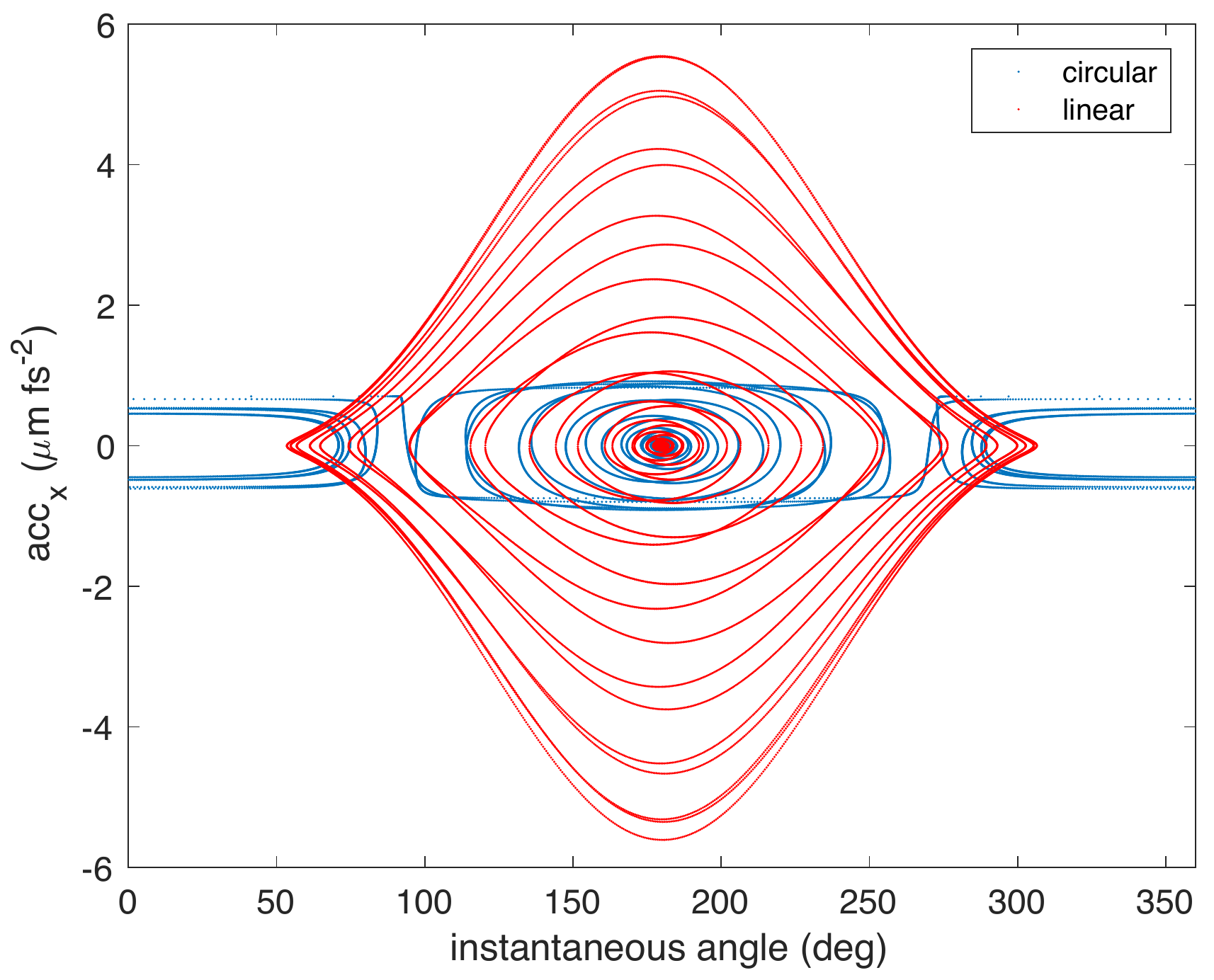}
\caption{Transverse acceleration ($x$-direction) as a function of the instantaneous angle of motion in the $x, z$-plane, $\arctan (u_x/u_z)$. The electron has an initial $\gamma_0=1000$  and collides with a 27fs (10-cycle) FWHM laser pulse of peak intensity $a_0=200$. Red lines: linear polarisation, blues lines: circular polarisation. \label{fig:angle_v_ax} }
\end{figure}

Figure \ref{fig:traj} shows the lab-frame trajectories for the electron in the baseline case for linear and circular polarisation. In this figure we also show the transverse ($x$-coordinate) accelerations as a function of time and longitudinal position. 
It is well known that a charged particle exhibits a figure-of-eight orbit in a linearly polarised field and a circular orbit in a circularly polarised one. However, due to the relativistic nature of these interactions, when observed from the lab frame the longitudinal components of the orbits become elongated and distorted. 
For both polarisations the pulse intensity is high enough relative to the $\gamma$-factor that the electron is reflected backwards during part of its interaction with the laser. Therefore we see some overlap in the particle path before the electron exits the pulse. Note that it is not until \emph{after} the electron has lost most of its energy (due to de-acceleration in the longitudinal $z$-direction) that it experiences significant acceleration in the transverse (i.e. $x$, $y$-) directions.
We see that in the case of linear polarisation the peak accelerations are confined to very short (sub-femtosecond) timescales. They are also much greater than the peak accelerations in the circular case. This results in a series of attosecond radiation flashes (two for each laser cycle), all in the same direction. While in theory this would be useful for determining the number of cycles in the pulse, any detector would be swamped by the $\gamma$-rays produced in the initial stages of the collision (when RR effects slow the electron down) and likely unable to resolve the rapid series of flashes occurring immediately afterwards.

What is less clear from Figure \ref{fig:traj} is how the acceleration is related to the change in angle for the two polarisation cases. In Figure \ref{fig:angle_v_ax} we plot the transverse acceleration ($x$-direction) as a function of the instantaneous angle of motion in the $x, z$-plane, $\arctan (u_x/u_z)$. In a circularly polarised field changes in acceleration occur over a very small angular range, meaning that the resulting radiation emissions will be confined to a very tight angle. (We also see that subsequent changes in acceleration occur at different angles meaning that the emissions from one laser cycle will be in a different location to the next.) Contrast this with the case of linear polarisation where we see changes in acceleration occurring over a much broader angular range.

\begin{figure*}
\includegraphics[width=0.49\textwidth,]{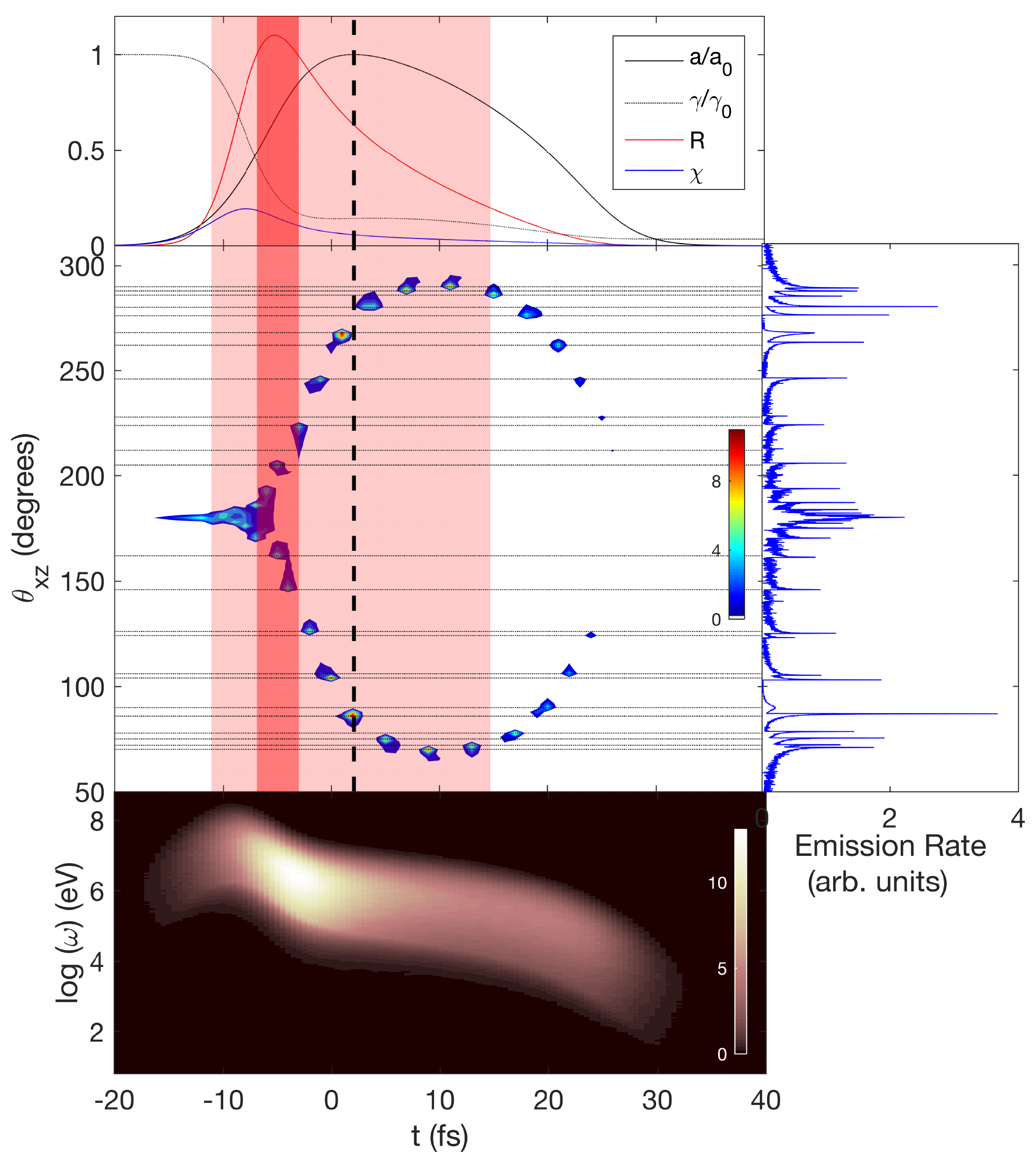}
\includegraphics[width=0.49\textwidth,]{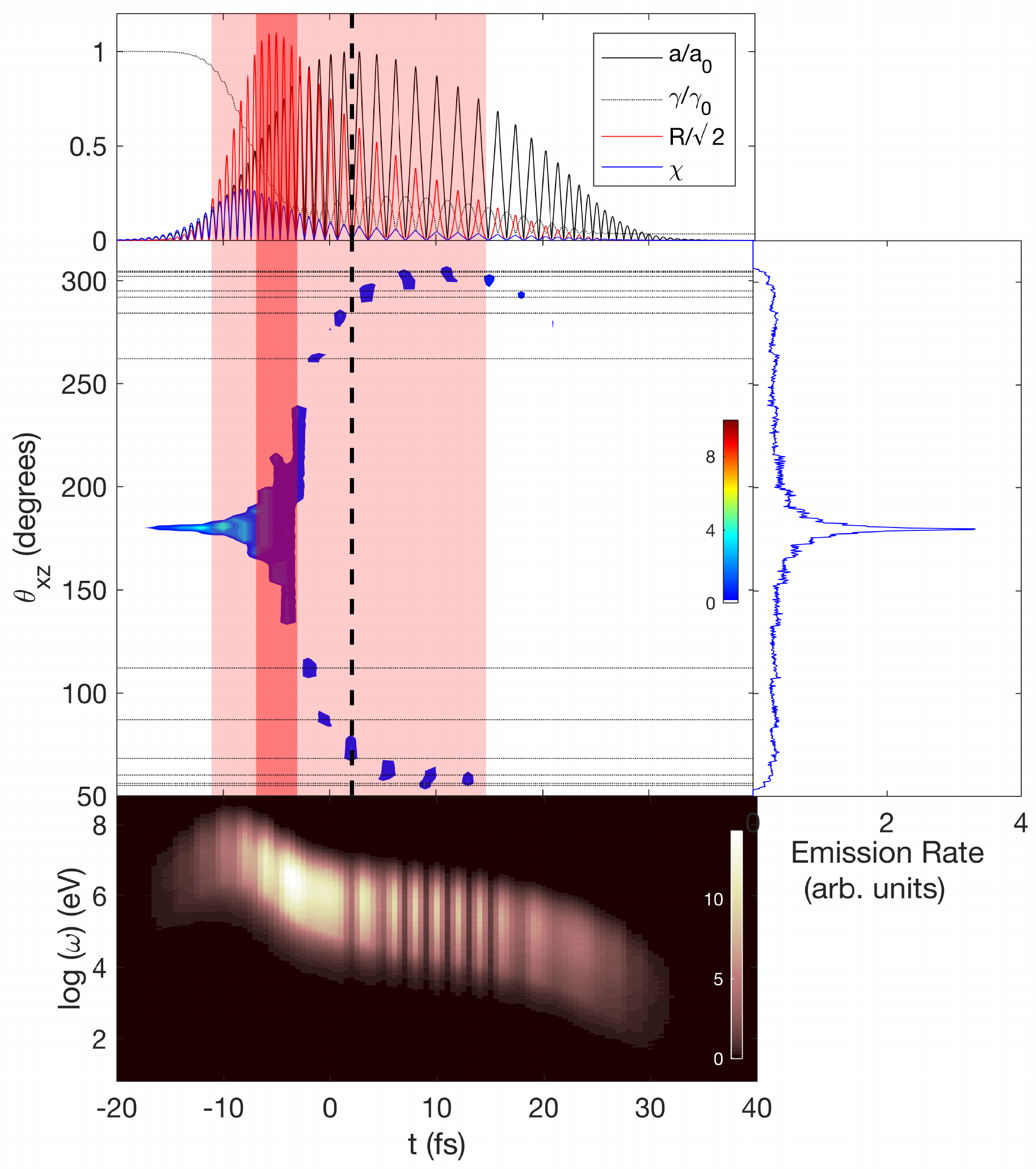}
\caption{The emission spectra and associated parameters for an electron of $\gamma_0=1000$ colliding with a 27fs (10-cycle) FWHM laser pulse of peak intensity $a_0=200$.  The left hand plots are for circular polarisation and the right hand plots linear. Top panels: normalised variables quantifying the interaction. Solid black lines show the intensity $a(t)$ of the field as experienced by the electron, normalised by the peak intensity $a_0$. Dotted black lines show the particle $\gamma$-factor normalised by the initial $\gamma_0$. Red lines: radiation reaction parameter $R(t)$. Blue lines: quantum efficiency parameter $\chi$. 
Centre panels: Angular distribution of emitted radiation as a function of time.  The horizontal black dotted lines mark the angles of peak emission for easy reference to the right hand panels. The vertical black dashed lines mark the time when the electron is in the most intense part of the field ($a=a_0$). Note that this may not overlap with the time of peak emissions since the $\gamma$-factor is already much lower by this point.
The regions shaded red in the centre and top panels demark the region where radiation reaction effects dominate (i.e. $R>1$). The regions shaded pink show where radiation reaction effects are important but non-dominant ($R>0.2$).
Right-hand panels: radiation emission rate as a function of angle (integrated over all frequencies).
Bottom panels: emitted rate as a function of time and frequency (integrated over all angles).
\label{fig:more_detail} }
\end{figure*}

To illustrate this more clearly, in Fig.~\ref{fig:more_detail} we plot details of a typical interaction. Once again the plots are for our baseline configuration, with the left hand set of plots showing circular polarisation and the right hand set linear. (For the purposes of this setup the electron is timed such that it would reach the peak of the laser at $t=0$fs  were it not to lose energy.) The top panels show the parameters that quantify the collision. From these we can see that the electron $\gamma$-factor rapidly decreases as soon as the electron enters the pulse, reducing to less than 20\% of its initial value before the electron reaches the peak field. This means that the product of $a$ and $\gamma$ is always much smaller than $m$ and so the quantum efficiency parameter $\chi$ remains low throughout the interaction. On the other hand the radiation parameter $R$ does become large, exceeding one at the start of the collision and remaining about 0.2 for most of the interaction. Hence RR effects play a crucial role, but we are in a predominantly classical regime rather than a quantum one.  The centre panels in Fig.~\ref{fig:more_detail} show the radiation emission rate as a function of time and angle (in the lab frame). In both cases there is an initial burst of radiation when the electron first enters the front tail of the laser pulse. At this point the collision is characterised by higher electron energy and lower field intensity meaning that the radiation is mostly in the forward direction (180 deg), see Ref.~\cite{PhysRevA.93.022112}. After this the characteristics for the two polarisation cases diverge. For the case of circular polarisation the emissions are largely confined to two narrow peaks for every laser cycle, at the top and bottom of the elliptical trajectory. For the case of linear polarisation the elongated figure of eight orbit means that the emissions are spread over a larger angle. This can be seen quite clearly by the fact that the peak emissions cover a continuous angular range between $t=-5$ and 0fs. Once $\gamma$ has reduced enough to take us out of the radiation dominated regime ($t>0$fs), but where RR effects are still important (i.e. $R\gtrsim 0.2$), we find that the emissions revert to isolated spikes at the two ends of each orbit. However, even then the figure-of-eight motion means that the radiation is not confined to such a tight angle as it is with circular polarisation (observe that the patches of blue don't contain the spots of red/intense emissions that we see in the plot for circular polarisation). This is very much evident in the (frequency) integrated \RefB{emissions} (right hand panels) where we only see small bumps corresponding to each of the main emission angles for linear polarisation, rather than the tall spikes seen for circular.  Away from the dominant spikes in emissions there remains a low level background at all angles which, when integrated, gives the non-zero background in the right hand panels. (Note that the lowest intensity emissions are coloured white in the centre panels to aid clarity.) Finally, the lower panels show the time evolution of the emitted rate per unit frequency, integrated over all angles. We can see that the initial burst of radiation when the electron enters the laser pulse reaches energies as high as $100$MeV, while the later attosecond bursts are in the range of 10's keV to 10's MeV. As expected from Figure \ref{fig:traj}, in the case of circular polarisation there are continuous emissions for the whole duration of the interaction, whereas for linear polarisation we see that the total emissions occur in bursts of <1fs duration.

\begin{figure}
\includegraphics[width=1.0\columnwidth,]{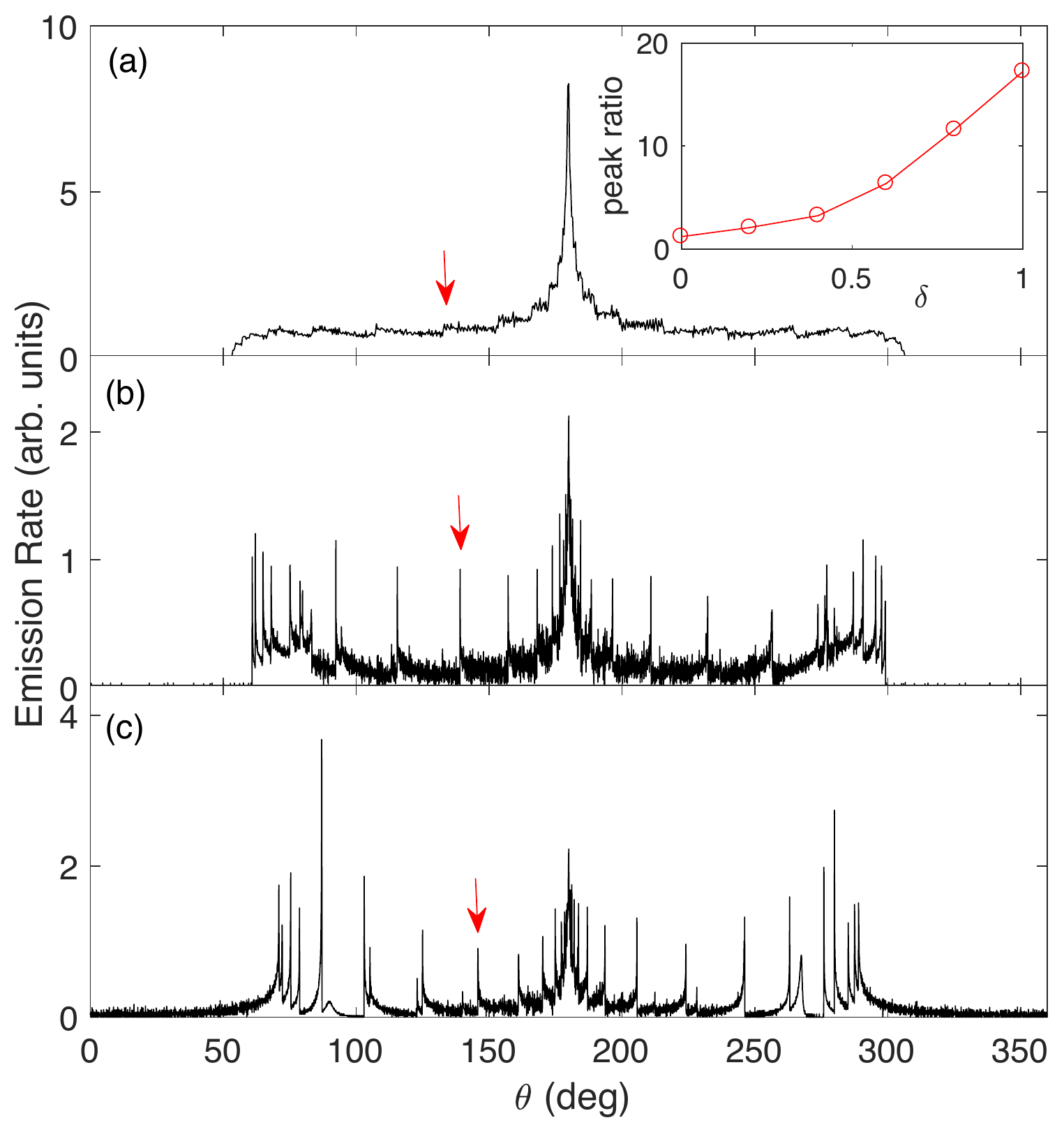}
\caption{\RefB{Angular radiation distribution} for the baseline example with varying polarisation: (a) linear $\delta=0$, (b) elliptical $\delta=0.5$, (c) circular $\delta=1$. The inset plot shows the ratio of the height of the peak (identified in the panels by red arrows) relative to the background for a number of different polarisations.\label{fig:polarisation} }
\end{figure}

In Figure \ref{fig:polarisation} we show the \RefB{angular emission rates} for three different polarisations: linear ($\alpha=0$), elliptical ($\alpha=0.5$) and circular ($\alpha=1$). From these plots we can see how the attosecond spikes emerge as the polarisation changes from linear to circular. In the inset plot we show the ratio of the height of a typical peak (identified by the red arrows) compared to the background, for a range of polarisations. (Note that the vertical axis scales are different in each of the three main panels.) We find  the spikes appear quite quickly as the polarisation factor $\delta$ increases, having amplitudes of several times the background for a field polarised to, e.g., $\delta=0.4$. Thus, by comparing the amplitudes of the measured peaks with the background radiation we would be able to deduce the degree of polarisation of the field.

\begin{figure}
\includegraphics[width=1.0\columnwidth,]{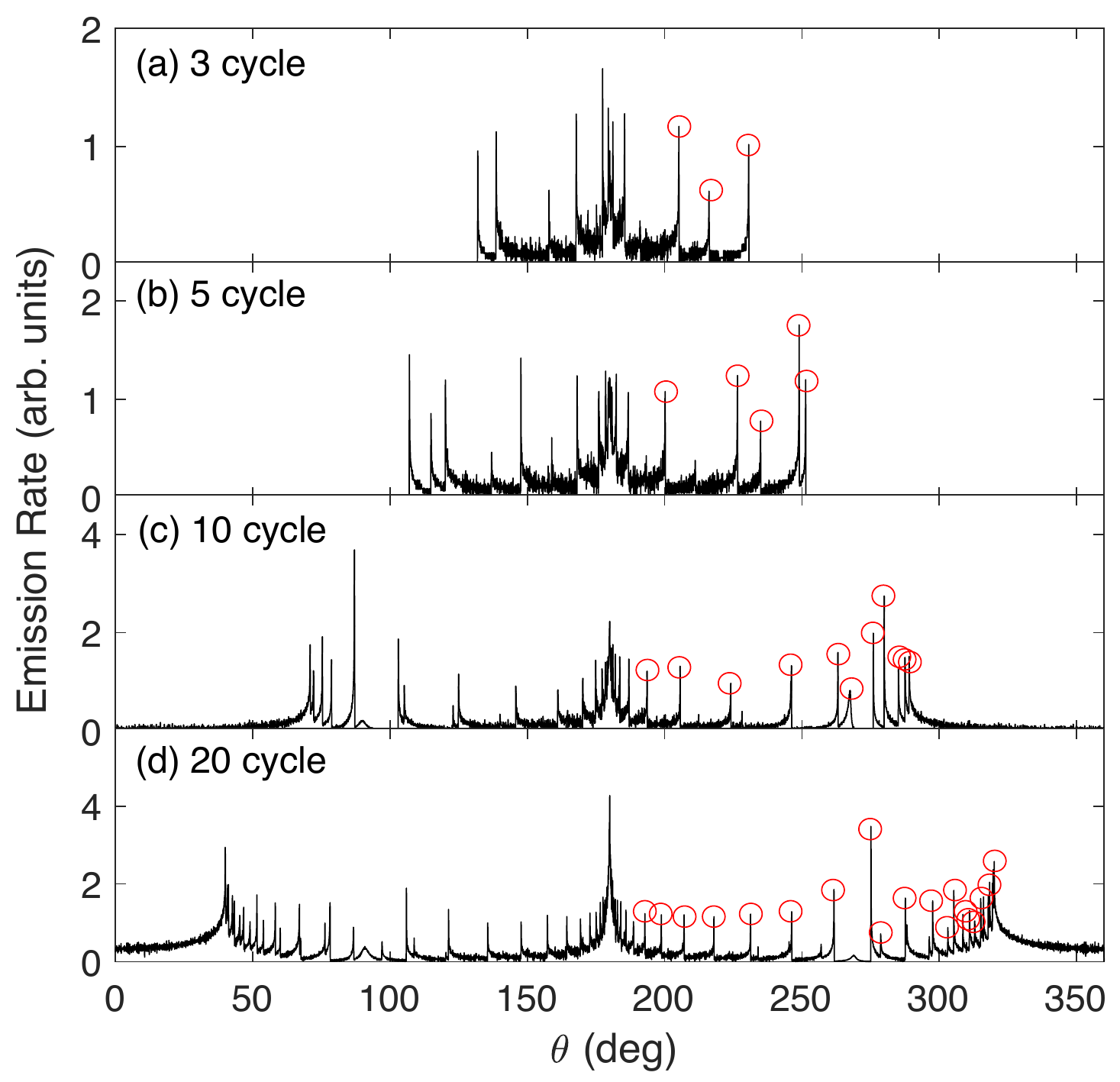}
\caption{Effect of changing pulse duration: (a) 3 cycles FWHM,  (b) 5 cycles, (c) 10 cycles, (d) 20 cycles. Apart from the number of laser cycles, the other parameters are as the baseline case and the laser is circularly polarised. \label{fig:duration} }
\end{figure}

Now we demonstrate the power of the results just presented. Figure \ref{fig:duration} shows \RefB{angular radiation distributions} from an electron in a laser pulse of four different durations. In the top panel the pulse is 8.1fs duration, equal to three cycles FWHM. Ignoring the broad bulk of emissions around $\theta=180^\circ$, we can count three peaks to the right of this \RefB{that have an amplitude more than twice the local background}. The next panel is for a 13.5fs, 5 cycle FWHM pulse. To the right of the broad emissions at $180^\circ$ we can count 5 peaks \RefB{whose amplitude is more than twice the neighbouring background}. Similarly for ten cycles we count eleven peaks, and for 20 cycles we count approximately 20 peaks. Thus with a $4\pi$ detector it would be possible to determine with good accuracy the number of cycles in the laser pulse.

\begin{figure}
\includegraphics[width=1.0\columnwidth,]{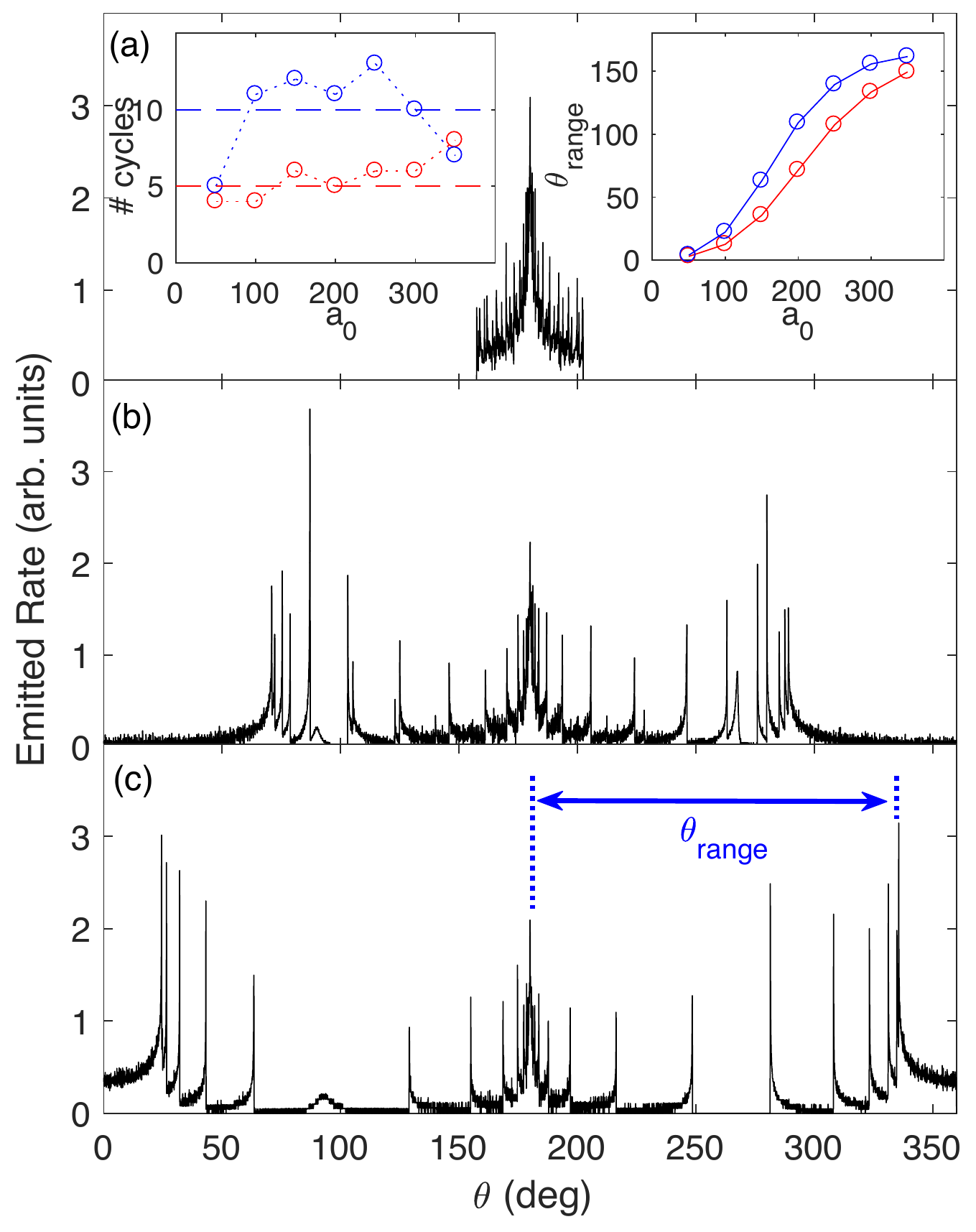}
\caption{Effect of changing peak intensity. Main panels show the \RefB{angular radiation distribution} for the baseline, circularly polarised, case but with (a) $a_0=100$,  (b) $a_0=200$, (c) $a_0=300$. The lefthand inset of panel (a) shows the counted number of cycles to the right of the broad peak at $\sim 180^\circ$ for a range of laser intensities and 5 and 10 cycles FWHM pulse duration. The righthand inset shows the angular range of the emissions, $\theta_{\textrm{range}}$ (defined in panel (c)), also for a range of intensities and 5 and 10 cycles duration. \label{fig:a0} }
\end{figure}

In Figure \ref{fig:a0} we show the effect of changing the peak intensity $a_0$. From the three main panels we see that the total angular range, $\theta_{\textrm{range}}$,  increases as $a_0$ increases. This is quantified in the top right inset of panel (a) which shows $\theta_{\textrm{range}}$ as a function of $a_0$ for pulses of two different durations. Thus, by measuring the angle of the furthermost peak we are able to determine the peak intensity of the field. In the lefthand inset of panel (a) we show the number of peaks in the distribution (to the right of the main bulge at $\sim 180^\circ$) as a function of $a_0$ for a 5 and 10 cycle pulse. The total number of spikes fluctuates slightly, but not too significantly, showing that the method of counting peaks is robust enough to give us a decent estimate of the number of cycles over a range of intensities. We find that the method works well over the range $a_0\in [50, 300]$ (i.e. from $2.1\times 10^{22}$ to $7.6\times 10^{23}$ W/cm$^2$). Below $a_0=50$ the weaker part of the field doesn't have enough strength to drive the attosecond emissions. Above $a_0=300$ the electron loses so much energy that it is reflected backwards before it reaches the most intense part of the field.

\begin{figure}
\includegraphics[width=1.0\columnwidth,]{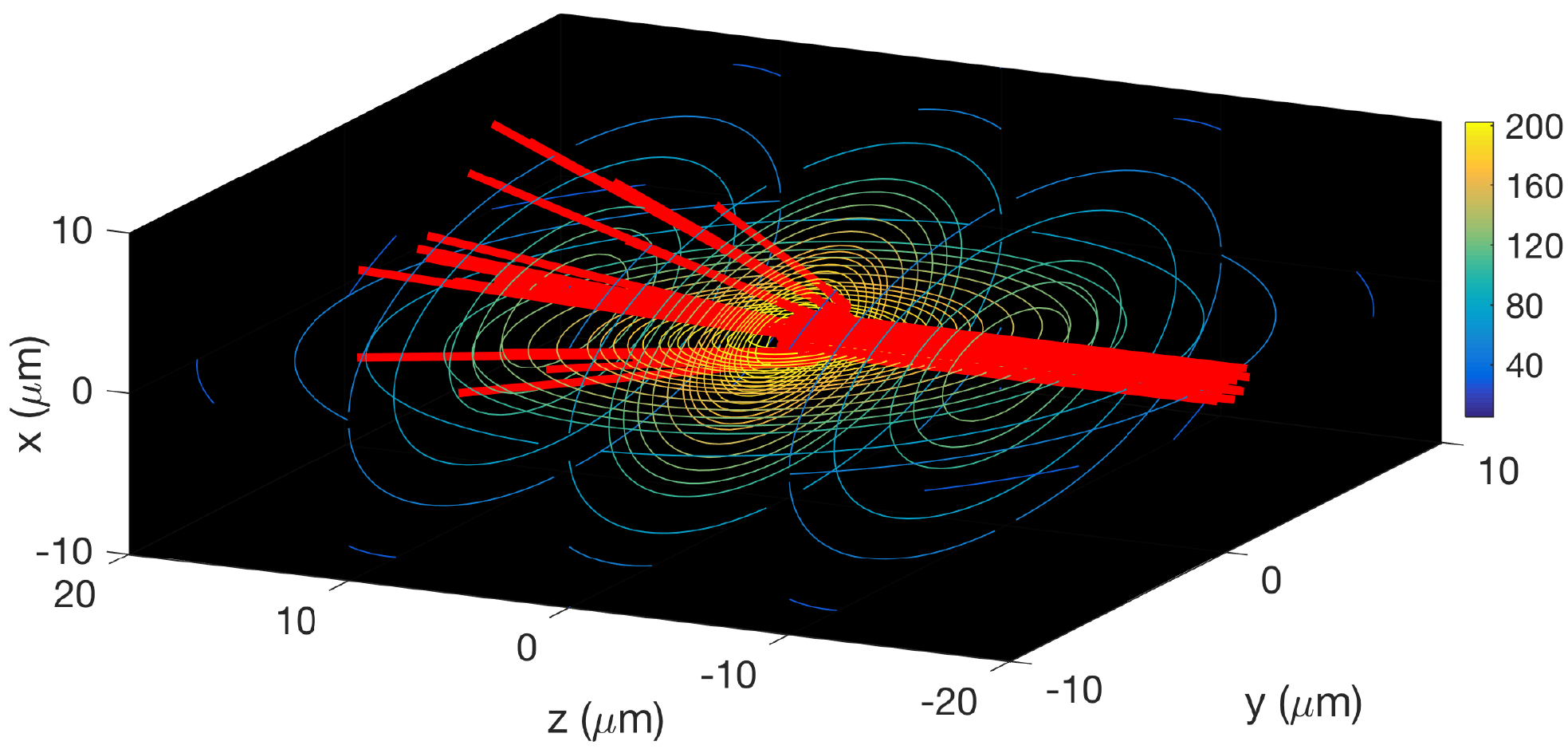}
\includegraphics[width=1.0\columnwidth,]{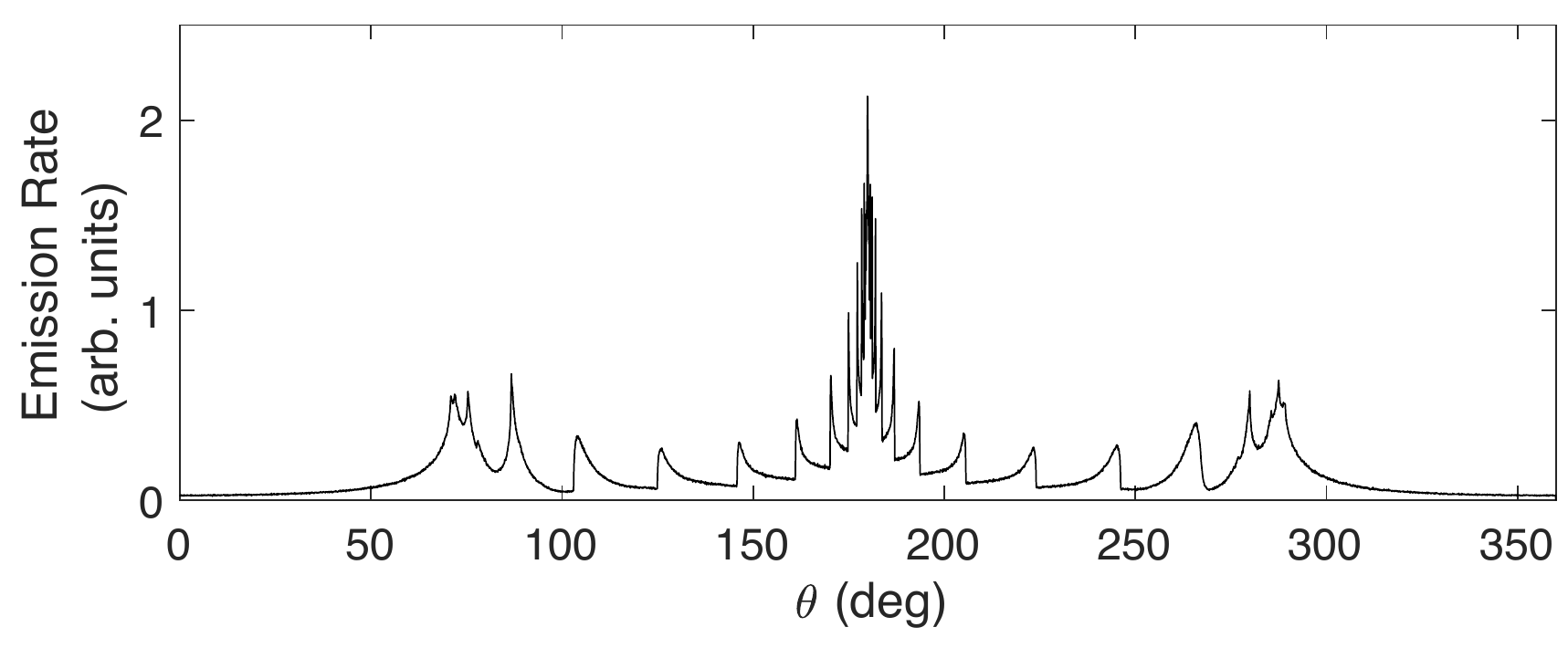}
\caption{Realistic example of an electron bunch colliding with a focussed laser pulse. The laser is modelled as a circularly polarised paraxial beam focussed to waist of $10\mu$m, with peak intensity $a_0=200$, wavelength $\lambda=800$nm and of 27fs FWHM duration. The  \RefB{electrons are modelled as a disk consisting of} $10^4$ particles distributed in space according to a Gaussian distribution of $1.5\mu$m FWHM in the transverse ($x$- and $y$-) directions, and a mean $\gamma$-factor of 1000 with a FWHM of 1. The top panel shows the intensity of the laser pulse at the time of maximum intensity together with the tracks of a random sample of 20 of the electrons. The bottom panel show the resulting \RefB{angular distribution of}  the combined emissions of all $10^4$ electrons.  \label{fig:paraxial} }
\end{figure}

Finally, in Figure \ref{fig:paraxial} we demonstrate that the results still hold when we move from a single electron in a plane wave to a bunch of electrons in a focussed field. For the electron beam we take, as a realistic example, the ELBE linear accelerator at the Forschungszentrum Dresden-Rossendorf in Germany \cite{ARNOLD200857}. We assume that the high-charge mode beam is accelerated to $\gamma=1000$ and the normalized transverse emittance of 2.5 mm mrad is preserved, so that we can assume a parallel incoming beam in the simulation. The beam is then focused to a FWHM diameter of 1.5 $\mu$m at the interaction point.  The energy spread of the electron bunch is taken to be $\gamma=1$ FWHM ($\Delta\gamma/\gamma_0 =10^{-3}$ is feasible at this facility \cite{PhysRevA.81.022125}). The laser is modelled as a circularly polarised focussed paraxial beam of waist $10\mu$m, peak intensity $a_0=200$, wavelength $\lambda=800$nm, and of 27fs FWHM duration. Some sample trajectories of the electrons are shown in the top panel of Figure \ref{fig:paraxial} and the total emissions for \RefB{an infinitesimal slice of the electron ensemble (comprising $10^4$ electrons)} is shown in the bottom panel. We find that the \RefB{angular radiation distribution} is qualitatively the same as in the idealised cases we have been considering. The spikes coming from each cycle of the laser field are still clearly distinguishable, although their bases are somewhat broadened. (This is due to that fact that electrons further from the central axis will see a field of lower intensity that those at the centre \cite{PhysRevAccelBeams.19.094701} and this, as we have seen in Figure \ref{fig:a0}, will effect the angle of emission.) We also note that we only see seven cycles in this plot, compared to 10 for the equivalent plane wave case. This is a result both of the focussed laser field decaying faster than it's plane wave cousin and it having a longitudinal field component which further slows the electron down. Neither of these points are detrimental to our analysis since one would scale the relationship between number of spikes and number of cycles according to the field being studied.

\section{Conclusion}
We have presented a method for determining the characteristics of an ultra-intense laser pulse. The method works by probing the pulse with a relativistic electron and detecting the resulting angular radiation emissions. Normally attention is focussed on the initial burst of high energy $\gamma$-rays produced when the electron first enters the field. However, we have shown the the more slowly moving electron after this event is buffeted around by the field, emitting high-energy X-rays in time with the rise and fall of the optical cycles. By measuring the count, amplitude and angles of these emissions we can determine with good accuracy the peak intensity, duration and polarisation of the ultra-intense laser pulse.

\RefA{Finally, we also note that these results suggest a concurrent measurement of both the angular distribution and frequency spectra of $\gamma$ radiation could provide us with information on the time evolutions of the electron energy and laser intensity during the interaction. (The concurrent measurement of angularly dependent high-energy photon spectra could be carried out with the differential filtering technique as demonstrated by Ref.~\cite{Jeon} or the CsI-array detector \cite{Cole, Poder}.) From these data, the time-resolved evolution for the energy loss per cycle and the quantum efficiency parameter could be obtained at a sub-femtosecond time resolution. }

\acknowledgments
This research was supported by the Swedish Research Council, grants \# 2010-3727 and 2012-3320. The author thanks Amol Holkundkar for useful discussions.

\appendix*
\section{Sensitivity to the initial electron energy}
\begin{figure}[b]
\includegraphics[width=1.0\columnwidth,]{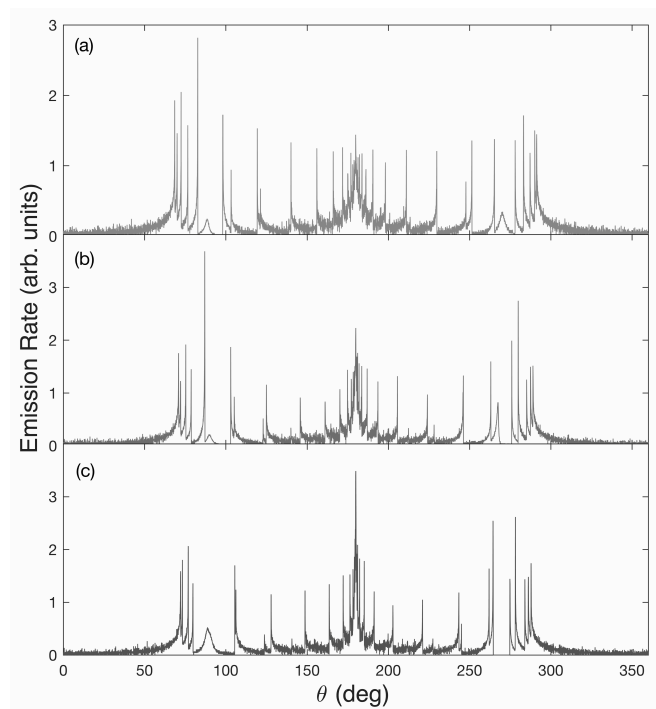}
\caption{\RefB{Effect of changing the initial electron energy: (a) $\gamma_0=500$,  (b) $\gamma_0=1000$, (c) $\gamma_0=2000$. Apart from the number of laser cycles, the other parameters are as the baseline case and the laser is circularly polarised. } \label{fig:gamma} }
\end{figure}

\RefB{In Figure \ref{fig:gamma} we show the emission rates for a selection of different electron $\gamma$-factors. It can be seen that the angular location of the spikes is relatively insensitive to the initial electron energy. The reason for this is that the angular spread only starts to occur once the electron has lost most of its energy to RR. Regardless of the energy before the collision, once the electron has radiated most of its energy and settled into a steady state the range of possible resulting energies is relatively small, see, e.g., Ref.~\cite{Yoffe} for a discussion. Thus, by the point in the collision where the angular emissions are radiated, the electron energy will fall within a narrow window regardless of its initial value. This means that the spikes will occur at roughly the same angles.}

%

\end{document}